\documentclass{aa}
\usepackage{placeins}
\usepackage{graphicx}
\usepackage{txfonts}
\usepackage{verbatim}
\usepackage{appendix}
\usepackage{hyperref}
\hypersetup{
    colorlinks=true,
    citecolor=blue,
    linkcolor=blue,
    urlcolor=blue,
}

\usepackage{color}
\usepackage{xcolor}
\definecolor{byzantine}{rgb}{0.74, 0.2, 0.64}
\definecolor{hotmagenta}{rgb}{1.0, 0.11, 0.81}
\newcommand{\valerio}[1]{\textcolor{black}{#1}}
\newcommand{\valentina}[1]{\textcolor{black}{#1}}
\newcommand{\ds}{$\delta$~Scuti}

\begin{document} 

   \title{TESS search for substellar companions \\ through pulsation timing of \ds{} stars}

   \subtitle{I. Discovery of companions around Chang~134 and V393~Car}

   \author{
   V.~Vaulato\thanks{E-mail: valentina.vaulato@unige.ch} \inst{1} \and
   V.~Nascimbeni\thanks{E-mail: valerio.nascimbeni@inaf.it} \inst{2} \and
   G.~Piotto\inst{1,2} 
          }

   \institute{Dipartimento di Fisica e Astronomia, Universit\`a degli Studi di Padova, Vicolo dell'Osservatorio 3, 35122 Padova, Italy \and INAF -- Osservatorio Astronomico di Padova, Vicolo dell'Osservatorio 5, 35122 Padova, Italy
             }

   \date{Accepted: 2 September 2022}

  \abstract{
  Early-type main-sequence pulsating stars such as \ds{} variables are one of the least explored class of targets in the search for exoplanets. Pulsation timing (PT) is an alternative technique to the most effective search methods. It exploits the light-travel-time effect (LTTE) to infer the presence of additional massive bodies around a pulsating star by measuring a periodic phase modulation of its signal. PT has been extremely successful in discovering and characterizing stellar binaries when it was applied to high-precision light curves over long temporal baselines, such as those delivered by the Kepler mission. In favorable conditions, the sensitivity of PT can reach the planetary-mass regime; one such candidate has already been claimed. The advent of TESS, with its nearly full-sky coverage and the availability of full-frame images, opens a great opportunity to expand this field of research. In this work, we present a pilot study aimed to understand the potential of PT as applied to TESS data, which are considerably different from Kepler data in terms of photometric noise, sampling cadence, and temporal baseline. We focused on the most favorable class of \ds{}, that is, those showing strong pulsations and very simple frequency spectra. After the development of a customized pipeline, we were able to detect candidate companions for two targets in the (sub-)stellar mass regime: Chang~134 ($43\pm 5$~$M_\mathrm{jup}$, $P\simeq 82$~d) and V393 Car ($\gtrsim 100$~$M_\mathrm{jup}$, $P\gtrsim 700$~d). Our results also highlight the limiting factors of this technique and the importance of an accurate absolute time calibration for future missions such as PLATO.
  }


   \maketitle
%
\section{Introduction}

Since the pioneering detections of 51~Peg~b,
the first exoplanet orbiting a solar-type star \citep{MAYOR&QUELOZ1995}, and HD209458b, the first transiting hot Jupiter \citep{Henry2000,Charbonneau2000}, contemporary exoplanetary research has mostly focused on 
 late-type (FGKM) main-sequence hosts. This is not just because of the interesting analogies with our own Solar System, but also because these stars are particularly suited for the  
most fruitful detection techniques 
in terms of discoveries: radial velocities (RV) and transits. These two techniques
are hampered by the presence of stellar activity and pulsations, making the detection and characterization of planets around very young, evolved, or early-type stars extremely challenging. Along these lines, the somewhat standard road to discovery has become 
 the photometric detection of transit-like signals, followed by an RV monitoring to confirm the planetary nature of the candidate and to measure its mass.

The vast majority of transiting planets known today have been discovered by dedicated space-based telescopes: starting with the pioneering work of the CoRoT satellite (2006-2013; built by an international consortium led by CNES; \citealt{Auvergne2009}), which was followed by the Kepler mission by NASA
(2009-2014; \citealt{Borucki2010}). Kepler was 
stopped because of a technical failure and
was then restored for a second-phase observing program called
K2 (2014-2018; \citealt{Howell2014}).
Finally, the advent of the NASA \textit{Transiting Exoplanet Survey Satellite} (TESS; \citealt{Ricker2015}), launched
in 2018 and currently operating, doubled the number of known 
candidate exoplanets by starting the first all-sky space-based transit survey. 
TESS is optimized to search for Earth- to Neptune-sized planets transiting bright and nearby main-sequence stars
with a particular focus on cool, red dwarfs (K and M) because their smaller radii and lower masses make the transit and RV signals much stronger. TESS objects of interest are three magnitudes brighter than those of Kepler on average \citep{Barclay2018}, enabling an easier and more effective
spectroscopic follow-up analysis. Most importantly in this context, and unlike its predecessors, TESS has the capability to download photometric data not only for a limited sample of preselected targets, but also for virtually any other bright star in the sky, although at a longer cadence with respect to the core sample. This is made possible by the availability of full-frame images (FFIs), originally downloaded every 30 minutes, and since sector 27, at a 10-minute cadence. FFI photometry opened a vast array of opportunities in many astrophysical fields, including extending the search for planets to a wider range in the stellar parameter space \citep{Montalto2020, Nardiello2019}.

Several alternative detection techniques have been proposed and implemented to search for exoplanets around stars other than solar-type or M dwarfs, or in different Galactic environments. A-type stars, in particular, are one of the least explored class of planet-hosting stars. Of 
more than $5\,000$ confirmed planets with known mass and/or radius listed in the NASA Exoplanet Archive, 
only 22 are hosted by 21 main-sequence stars earlier than the F0V spectral type ($T_\mathrm{eff}\geq 7220$~K; \citealt{Pecaut2013}). With only four exceptions coming from direct imaging surveys, they were all detected through the transit technique (3 by HAT-Net, 3 by WASP, 5 by KELT, 2 by MASCARA, 3 by Kepler, 2 by TESS) and were confirmed by means of spectral tomography \citep{CollierCameron2010} and/or statistical validation techniques because classical RV measurements are usually not effective on such stars, whose spectral lines are few and Doppler broadened by their very high rotational velocities ($v\sin i \sim 100$~km/s; \citealt{Rainer2021}).

Only three of the hosts described above are confirmed or suspected \ds{} variables, that is, population I pulsators located in the classical Cepheid instability strip (\citealt{Bowman2018,Guzik2021} and references therein): WASP-33b \citep{Herrero2011}, HAT-P-57b \citep{Hartman2015}, and WASP-167b = KELT-13b \citep{Temple2017}. Because $\sim 50\%$ of the stars in this region of the parameter space are expected to show \ds{} pulsations to some extent \citep{Bowman2017}, some astrophysical reason must explain the lack of planets around \ds{} stars, and/or a strong observational selection effect is at play. The latter explanation is likely true, as the stellar pulsation pattern is known to limit the efficiency of transit-search algorithms even after they are filtered in the frequency domain by sophisticated techniques \citep{Ahlers2019,Hey2021}.

In this context, the pulsation timing technique (PT; \citealt{Hermes2018}) is a completely independent method for detecting massive companions around stars showing coherent and well-detectable pulsations, such as \ds{} stars, pulsating white dwarfs, or hot subdwarfs. The underlying physical principle is the so-called light-travel-time effect (LTTE) that was first discussed in detail by \citet{Irwin1952}: Because the speed of light $c$ is finite, the motion of the target along the line of sight will translate into a phase shift of any periodic astrophysical signal originating from the star, including pulsations. The arrival times of the signals can then be compared with those predicted by a linear ephemeris (implying a constant periodicity $\mathrm{d}P/\mathrm{d}t = 0$) by computing the so-called O$-$C diagram (observed $-$ calculated; \citealt{Sterken2005}). Any deviation from a strict periodicity $\mathrm{d}P/\mathrm{d}t \neq 0$ (assuming that the pulsation themselves are intrinsically stable) would then reveal itself as a nonzero second derivative in the O$-$C diagram (see Sec.~\ref{Section:Harmonic_analysis} for details). The LTTE signal from a single, massive perturber on a Keplerian orbit is formally analogous to an RV curve having half of the eccentricity $e$ and the argument of the periapsis $\omega$ decreased by $\pi /2$ \citep{Irwin1952}; if the full shape of the LTTE signal is retrieved, then the orbital period $P$, the minimum mass $M\sin i,$ and other orbital elements of the perturber can be reliably measured. The PT approach shares some similarities with the astrometric method \citep{Black1982} in that it measures the displacement of the target star with respect to an inertial rest frame, but projected along the line of sight rather than on the sky plane. Notably, also the underlying selection effects are similar because both methods are more sensitive to massive bodies at large star-planet separations. PT therefore probes a region of the parameter space that is complementary to the space that is investigated by transits and RVs, which are more sensitive to close-in perturbers \citep{Sozzetti2005}.

The PT technique requires photometric data spanning a temporal baseline that is long enough to sample the orbital period of the perturber and a signal-to-noise ratio (S/N) high enough to constrain the phase shift of the oscillations to a level comparable to the expected LTTE amplitude\footnote{As a comparison, the LTTE effect induced by the orbital motion of Jupiter around the Sun, as seen edge-on, is $\sim 10$~s \citep{Schneider2005}, with a period of $\sim 12$~years. See also Sec.~\ref{Section:Discussion_and_conclusions}.}. The main pulsation modes of the star have to be individually identified and measured on the Fourier spectrum of the light curve, implying that continuous time series spanning at least $\sim 10$~days are mandatory to achieve the needed resolution in the frequency space ($\sim 0.1$~cycles/day) when dealing with the harmonic content of typical \ds{} stars \citep{Murphy2014}. This is the main reason why sparse sampling imposed by the day-night cycle is the main limiting factor of ground-based observations.  
\valerio{Targets with an extremely simple harmonic content, such as pseudo-sinusoidal pulsators, represent an exception, and a few pioneering results were published by analyzing ground-based data \citep{Paparo1988,Barlow2011a}. Interestingly, the latter claim was subsequently confirmed by an independent RV follow-up, demonstrating the reliability of this technique \citep{Barlow2011b}.}

\valerio{The full power of the PT technique, however, revealed itself when} the nearly uninterrupted 4 yr baseline of Kepler photometry was exploited. Some works \citep{Shibahashi2012,Murphy2014,Shibahashi2015,Murphy2015,Murphy2018,Murphy2020}, although they focused on the detection and characterization of stellar-mass companions, demonstrated that under favorable assumptions, even LTTE signals from planetary-mass companions ($\lesssim 13$~$M_\mathrm{jup}$) are in principle detectable through space-based photometry \citep{Murphy2014}.  \cite{Murphy2016} presented the PT detection of a massive planetary candidate for the first time ($11.8\pm 0.7$~$M_\mathrm{jup}$, $P=840\pm 20$~d). This candidate orbits a metal-poor \ds{} star (KIC~7917485). Together with the discovery by \cite{Silvotti2007} of a giant planet around the hot subdwarf B star V391~Peg, these are the only PT discoveries of planetary-mass objects 
published so far. \valerio{It should be noted, however, that a follow-up paper on V391~Peg with new data \citep{Silvotti2018} failed to fully reproduce the previous claim, probably due to nonlinear interactions between pulsation modes (see also \citealt{Bowman2021}). This acts as a caution.}

While several studies of \ds{} stars have been published that exploited TESS short-cadence data (\citealt{Antoci2019}; \citealt{Hasanzadeh2021}; \citealt{Southworth2021} for a review), none of them applied the PT technique. The goal of this paper is to focus on the PT analysis of a small sample of particularly suited targets observed by TESS at short cadence as a pilot study for the exploitation of TESS (and, later on, PLATO) data on a larger scale. In Section~\ref{Section:Sample_selection} we describe how our targets were identified and how their relevant stellar parameters were collected. In Section~\ref{Section:Harmonic_analysis} we summarize all the steps of our data analysis, starting from the filtering of raw light curves through the harmonic analysis to measure the phase shifts as a function of time. We determine the best-fit orbital solutions through an LTTE model in Section~\ref{Section:Orbital_solutions} and discuss the results in Section~\ref{Section:Discussion_and_conclusions}.

\section{Target selection and characterization}
\label{Section:Sample_selection}

\begin{table*}[h!]\centering
      \caption[]{Adopted stellar parameters and other basic information for Chang~134 (second and third columns) and~V393 Car (fourth and fifth columns). See Section~\ref{Section:Sample_selection} for details.}
         \begin{tabular}{l|cl|cl}
            \multicolumn{1}{c}{} & \multicolumn{2}{c}{\textbf{Chang 134}} & \multicolumn{2}{c}{\textbf{V393 Car}} \\
            \hline
            parameter   & value & source (\& ref.)& value & source (\& ref.)\\
            \hline\hline
            TIC (ID) & 431589510 &  TICv8 (A) & 364399376 & TICv8 (A) \rule{0pt}{15pt}\\
            TYC (ID) & 9158-919-1 & Tycho-2 (B) & 8911-2754-1 & Tycho-2 (B) \\
            HD (ID) & --- & --- & 66260 & Simbad (C)\\
            2MASS (ID) & J03224585-7237459 & 2MASS (D) & J07590267-6135009 & 2MASS (D)  \\
            R.~A.~[deg] & 50.69114 & Gaia EDR3 (E) & 119.76113 & Gaia EDR3 (E) \rule{0pt}{15pt} \\
            declination [deg] & $-$72.62943 & Gaia EDR3 (E)& $-$61.58361 & Gaia EDR3 (E)\\
            parallax [mas] & $0.75\pm 0.11$ & Gaia EDR3 (E) &  $5.00\pm 0.10$ & Gaia EDR3 (E)\\
            distance [pc] &  $1090\pm 120$  & Gaia EDR3 (F) &  $199 \pm 5$  & Gaia EDR3 (F) \\
            $T$ (mag) &  $11.89\pm 0.01$  & TICv8 (A) &  $7.185\pm 0.007$  & TICv8 (A) \\
            $B$ (mag) &  $12.35\pm 0.11$  &  Tycho-2 (B)&  $7.76\pm 0.01$ &  Tycho-2 (B) \\  
            $V$ (mag) &  $12.33\pm 0.17$   & Tycho-2 (B)&  $7.46\pm 0.01$ & Tycho-2 (B) \\
            $J$ (mag) &  $11.40\pm 0.02$  & 2MASS (D) &  $6.84\pm 0.02$ &  2MASS (D)\\ 
            $H$ (mag) &  $11.29\pm 0.02$  & 2MASS (D) &  $6.75\pm 0.06$ & 2MASS (D) \\
            $K_s$ (mag) &  $11.23\pm 0.02$  & 2MASS (D) &  $6.68\pm 0.02$ &  2MASS (D)\\
            $T_\mathrm{eff}$ [K]  & $7200\pm 300$ & seismic (G) & $7400\pm 200$ & seismic (G) \rule{0pt}{15pt}\\
             " & $7281\pm 170$ & Gaia DR2 (H) & $7040\pm 280$ & Gaia DR2 (H)\\
            $L_\star/L_\odot$ & ---& --- & 27.44 $\pm$ 0.32 & Gaia DR2 (H)\\
            $M_\star/M_\odot$  & $1.38\pm 0.03$ & \texttt{StarHorse} (I) & $2.02\pm 0.10$ & \texttt{StarHorse} (J) \\
            $R_\star/R_\odot$  & --- & --- & $3.52\pm 0.29$  &  Gaia DR2 (H)                           \\
            $\log g_\star$ [cgs]  &  4.04  & seismic (G) &   3.58   &  seismic (G)\\
 \hline
         \end{tabular}\label{tab:stellar_parameters}
\tablefoot{References: A) TESS Input Catalog Version 8 \citep{Stassun2019}; B) Tycho-2 \citep{Hog2000}; C) Simbad data base, \citep{Wenger2000}; D) Two Micron All Sky Survey (2MASS, \citealt{Cutri2003}); E) Gaia Early Data Release 3 (EDR3, \citealt{Gaia2021}); F) \citet{BailerJones2021}; G) \citet{BarceloForteza2020}; H) Gaia Data Release 2 (DR2, \citealt{GaiaCollab2018}); I) \citet{Queiroz2020}; J) \citet{Anders2022}. }
   \end{table*}

The vast majority of targets observed by TESS at short and ultra-short photometric cadence (120 and 20~s, respectively) are FGKM dwarfs selected for the core transit-search survey, drawn from the so-called candidate target list (CTL; \citealt{Fausnaugh2021}). \valentina{However, stars of different spectral types, including pulsating variables, are ingested mostly from GI/DDT proposals or from the TESS Asteroseismic Science Operation Center (TASOC\footnote{\url{https://tasoc.dk}}), which provides the TESS Asteroseismic Science Consortium (TASC) with an  asteroseismological database of the mission \citep{Schofield2019}.}

In order to select our targets, we chose the catalog compiled by \cite{Chang_2013} as a starting point. This catalog lists a sample of 1,578 \ds{} stars that were identified at high confidence level by several previous surveys.
We cross-matched this catalog with the TESS CTL tables for sectors from 1 to 42 included   
to obtain only \ds{} stars that were observed by TESS at two-minute cadence. 
As an additional constraint, we further restricted our sample by 1) requiring a minimum data coverage of at least seven TESS sectors, not necessarily contiguous, to build $O$-$C$ diagrams with a number of points and a temporal baseline long enough to detect LTTE signals at an $\text{approximately }$one-year timescale (see Section~\ref{Section:Orbital_solutions}); 2) selecting targets brighter than $T=12$ in the TESS photometric system, to avoid being limited by photon and background noise; and 3) excluding all the stars that have been identified as binaries in the \citet{Chang_2013} catalog. These additional constraints are justified by the nature of this study, which is not focused on a complete sample, but is rather intended as a pilot study to investigate the limiting factors of TESS (especially due the systematic errors) on a very small sample of the most favorable targets.

The final output of this selection process is a short list of 12 targets, which were then individually examined both through a literature search and by inspecting their TESS light curves. In particular, we confirmed that all of them are actually \ds{} pulsators and carried out a preliminary harmonic analysis on them by computing \valentina{the generalized Lomb-Scargle periodogram (GLS; \citealt{Zechmeister2018})} 
of a single sector with the same algorithms as described in Section~\ref{Section:Harmonic_analysis}. While most targets are classical \ds{} stars showing small-amplitude ($\sim$0.01-0.03~mag) pulsations and a complex pattern of different modes in the frequency domain, two of them stand out as high-amplitude pulsators ($\geq 0.1$~mag) with a particularly clean periodogram, in which most of the signal is due to a single, well-defined pulsation mode and its harmonics (Fig.~\ref{fig:lc_chang134}, \ref{fig:lc_v393}): hereafter we refer to them as Chang~134 (from its ID number in the \citealt{Chang_2013} catalog) and V393 Carinae. This combination of a large pulsation amplitude and a very simple and coherent frequency pattern makes these targets well suited for a precise timing analysis, 
especially for our pilot study. For this reason, we focused this analysis on them \valerio{and set aside the other targets (five of which are multi-mode pulsators, but with modes that are easy to identify) for the next paper of the series}. A short review of the available scientific literature is given in the following subsections, and the basic astrophysical parameters of the two targets are reported in Table~\ref{tab:stellar_parameters}.
It is worth mentioning that both our targets could fit the high-amplitude \ds{} subclass (HADS; \citealt{Breger2000,Antoci2019}), which is characterized by large-amplitude pulsations, low rotational velocities, and time-domain spectra that are dominated by radial modes. We did not attempt to perform a detailed identification of their pulsation modes, however, because our work is focused on a dynamical technique (where the pulsation pattern is exploited just as a coherent astrophysical clock) rather than on stellar physics. 


\subsection{Chang~134 = TYC~9158-919-1}
Chang~134, also known as TYC~9158-919-1 or TIC~431589510,  was first identified as a variable star by ASAS \citep{Pojmanski2002} and classified as a generic \ds{} star (DSCT variability class). No targeted follow-up study on it has been published since 2002. 
An asteroseismological measurement of $\log (g)$ and $T_\mathrm{eff}$ was included in the catalog by \citet{BarceloForteza2020} through the $\nu_{\max}$ seismic index extracted from the TESS light curves in an automated fashion. All the astrophysical parameters we collected are reported in the second and third columns of Table~\ref{tab:stellar_parameters}.

The PT technique, just like astrometry or RVs, cannot directly infer the mass of perturbing body, but rather its ratio $m_p/M_\star$ with the stellar mass. The latter is needed at the analysis stage (Section~\ref{Section:Orbital_solutions}) to properly interpret our results for the two targets. For Chang~134, an estimate of stellar mass ($M_\star = 1.40\pm 0.16$~$M_\sun$) and age was first published in the catalog by \citet{Mints2017} using UniDAM models, but without taking advantage of spectroscopic data or Gaia parallaxes, which were not available at that time. 
Later, Chang~134 was included in the large-scale analysis by \citet{Queiroz2020}, combining high-resolution spectra from the APOGEE-2 survey DR16 with broadband photometric data taken from different sources and Gaia DR2 distances. Stellar parameters are derived as the posterior distribution returned by the Bayesian isochrone-fitting code \texttt{StarHorse} \citep{Queiroz2018}. 
The stellar mass reported for Chang~134 is $M_\star = 1.38\pm 0.03$~$M_\odot$, where the central value is the $50th^\mathrm{}$ percentile of the \texttt{StarHorse} posterior distribution, while the symmetrized uncertainty is calculated as the 
half-difference between the $\mathrm{84th^{}}$ and the $\mathrm{16th^{}}$ percentiles of the same distribution.

\subsection{V393~Carinae = HD~66260}
  Unlike Chang~134, the much brighter V393~Car (also known as HD~66260 or TIC~364399376) has been the subject of several targeted studies. It was first discovered as a variable star and classified as a \ds{} radial pulsator by \citet{Helt1984}. \citet{Balona1999} found in a more detailed analysis that the dominant mode is rather nonradial; the debate of this point is still ongoing. A first claim of change in the period of the main pulsation mode appeared in \cite{Garcia2001}, who measured a 7.18-minute discrepancy with respect to the ephemeris by \citet{Helt1984}. Interestingly, further ground-based photometric follow-up by \citet{Axelsen2014} retrieved the very same period of \citet{Helt1984} within 0.33~s, and found no conclusive evidence of additional overtones. The present work therefore also represents an opportunity to confirm or disprove the claimed period change, and to determine whether it is caused by a secular or an oscillating term. 

A measurement of $\nu_\textrm{max}$, $\log (g)$, $T_\mathrm{eff}$ was listed by  for Chang 134 by \citet{BarceloForteza2020}. Unfortunately, V393~Car is missing from the \citet{Queiroz2020} catalog, but is listed in the \citet{Anders2022} catalog, which applied \texttt{StarHorse} on the Gaia~EDR3 photometric and astrometric measurements, yielding $M_\star = 2.02\pm 0.10$~$M_\sun$ for this target. 
As accurate stellar parameters of V393 Car derived from spectroscopy are lacking in the literature, we attempted to obtain an independent estimate of the stellar mass as a crosscheck by applying the empirical relations derived by \cite{Moya2018} as a function of other stellar parameters. In particular, we exploited the relation
\begin{equation}
    \label{eq:Moya_relation}
    \log_{10}(M_\star) = - a + b\, T_{\mathrm{eff}} + c \log_{10}(L_\star) ,
\end{equation}
where $a=-0.119\pm0.003$, $b=2.14\times10^{-5}\pm5\times10^{-7}$ and $c=0.1837\pm0.0011$ are the coefficients calculated by \cite{Moya2018}. This relation is valid within the range of temperatures 4780 $\leq T{_\mathrm{eff}} \leq$ 10990~K. Adopting $T_{\text{eff}}$ from the seismic value given by \citet{BarceloForteza2020} and $L_\star$ from the Gaia DR2 catalog\footnote{\valerio{Unfortunately, the stellar parameters reported for V393~Car by Gaia DR3 \citep{GaiaDR3} are highly discrepant with each other and associated with unreasonably small errors, possibly because the pipeline is unable to deal with a pulsating star. We therefore adopted the DR2 values for consistency.}}, and propagating the errors, we computed a stellar mass $M_\star = 1.98 \pm 0.17 M_{\odot}$ for V393 Car, which is perfectly consistent with the previous value. We therefore adopt the $M_\star = 2.02\pm 0.10$~$M_\sun$ estimate from \citet{Anders2022} in the subsequent analysis. 

We are aware that a detailed asteroseismic analysis performed on all the available TESS light curves would likely yield much more accurate fundamental parameters for both our targets \citep{Hasanzadeh2021}.  However, since $\text{an  error of about } 5\%$  on stellar mass will not be the dominant source of uncertainty in our final parameters (as we discuss in Section~\ref{Section:Orbital_solutions}), an analysis like this is beyond the scope of our work.

\subsection{TESS light curves}
Chang 134 and V393 Car were both observed by TESS in two-minute cadence mode. At the time when our analysis started, Chang~134 was observed in 9 nonconsecutive sectors (1-2-3-6-13-27-28-29-36) between July 2018 and March 2021. V393 Car was observed in 14 nonconsecutive sectors (1-4-7-8-10-11-27-28-31-34-35-36-37-38) between July 2018 and April 2021. All the photometric data analyzed in this work were processed and extracted from the raw data by the Science Processing Operations Center (SPOC) pipeline (\citealt{Jenkins2016}) and are publicly available on the Mikulski Archive for Space Telescopes (MAST\footnote{\url{https://mast.stsci.edu/portal/Mashup/Clients/Mast/Portal.html}}). Specifically, we built the light curves for our analysis by extracting the stellar flux from the pre-search data conditioning simple aperture photometry column (PDCSAP; \citealt{Smith2012,Stumpe2012}) because it results in cleaner time series since systematic long-term trends are removed.

\section{Harmonic analysis}
\label{Section:Harmonic_analysis}

Our approach to data analysis is based on the fit of sums of harmonic functions to preselected segments of our light curves in the form
\begin{equation}\label{eq:harmser}
    \sum_i f(A_i,P_i,\phi_i) = \sum_i A_i \times \cos\left (\frac{2\pi t}{P_i} + 2\pi\phi_i \right ) \textrm{ ,}
\end{equation}
where $A_i$ is the amplitude, $P_i$ the period, and $\phi_i$ the phase of each component normalized between 0 and 1.
The phase of the signal at the dominant pulsation mode $\phi_0$ is then retrieved for each segment, giving the so-called phase shift $\phi_0(t)$ as a function of time. From this quantity, the usual $O-C$ diagram in time units can be computed in a straightforward way by just multiplying the phase shift by the pulsation period $P_0$.

We developed an independent pipeline to perform all the needed tasks from the raw light curves down to the final $O-C$ diagram. Its flow chart (described in the following sections) is conceptually similar to the phase modulation (PM) method described by \citet{Murphy2014}, with some substantial differences, in particular, in how the actual fit is performed, that is, through a Markov chain Monte Carlo (MCMC) analysis in our case rather than with a simple least-squares algorithm.
Most of our processing steps were implemented by scripting and modifying routines from the \texttt{VARTOOLS} code version 1.39\footnote{\url{https://www.astro.princeton.edu/~jhartman/vartools.html}} released by \cite{Hartman_and_Bakos_2016}, to which we refer  
for a more in-depth description of the individual algorithms.

\subsection{Light curve preconditioning}
We started our process by downloading 
TESS SPOC light curves of our targets from the MAST archive for each observing sector (23 combined sectors). Each data point was sampled every two minutes as a default for  
short-cadence observations. 

As a first step, we discarded all 
data points flagged as bad, that is, with a quality factor $q\neq0$ (\texttt{QUALITY} column).  We then extracted the time column \texttt{TIME} from each FITS file and converted it into the $\textrm{BJD}_\textrm{TDB}$ standard \citep{Eastman2010}. Finally, we extracted the \texttt{PDCSAP\_FLUX} column and its associated error \texttt{PDCSAP\_FLUX\_ERR} and converted them into the magnitude system. The choice of PDCSAP over SAP is justified by our need that systematic errors, especially those manifesting themselves as long-term trends, are corrected for as much as possible to avoid unnecessary noise on our periodograms. As our analysis is focused on retrieving the phase of the pulsation signal, we are not concerned by any small perturbation in the amplitudes that might be introduced by the PDCSAP processing \citep{Cui2019}. In order to filter out the most obvious outliers that survived the $q=0$ selection, we carried out an iterative clipping at 15~$\sigma$ with respect to the mean. This specific threshold was empirically set by confirming that the shape of the main pulsation mode was left unchanged. 

The final step of our preconditioning recipe was to split the light curves from each sector into smaller chunks that were individually analyzed later to become single points in the $O-C$ diagram. Working on continuous segments, about$\text{ }$10~days has been shown to be a reasonable compromise between the need of getting 1) a frequency resolution that is high enough to reliably measure the phases of the individual pulsation modes and 2) a  time resolution on the $O-C$ diagram as low as possible \citep{Murphy2014}. Because for most TESS sectors the only significant interruption is the one-day central gap to allow the data downlink toward Earth at each perigee of the spacecraft, we chose that gap as a natural boundary and split each sector accordingly into two mostly continuous segments of $\text{about }$14~days each, to which we refer hereafter as ``orbits''. We are therefore left with 18 orbits for Chang~134 and 28 orbits for V393~Car. Each orbit was individually normalized to unit flux (and magnitude zero) before we continued to the next stages.

\begin{figure*}
    \centering
    \includegraphics[width=0.95\textwidth]{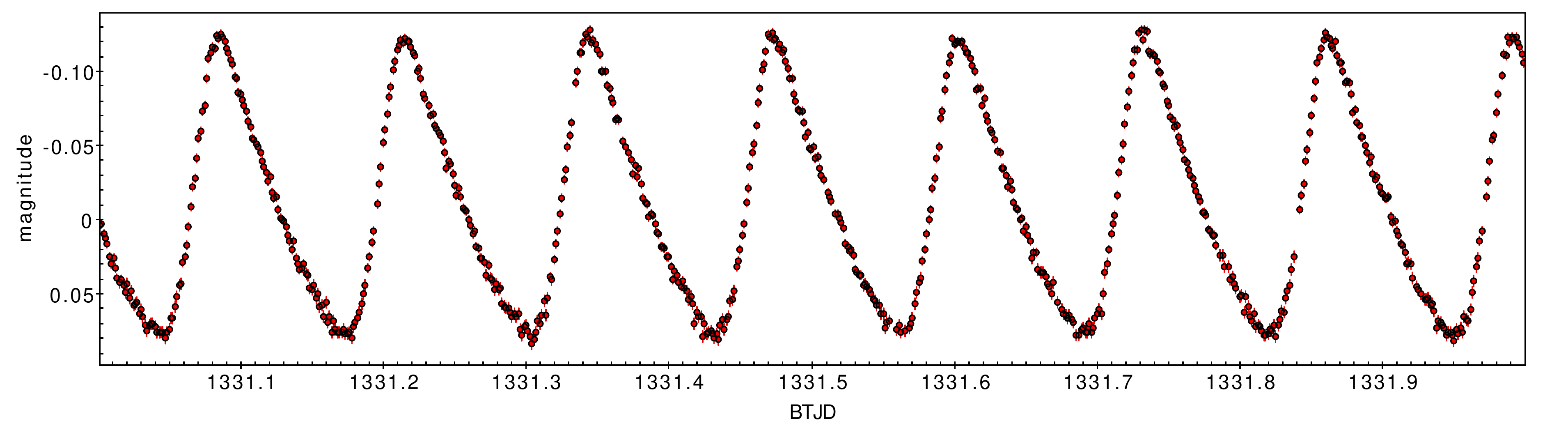}

    \includegraphics[width=0.95\textwidth]{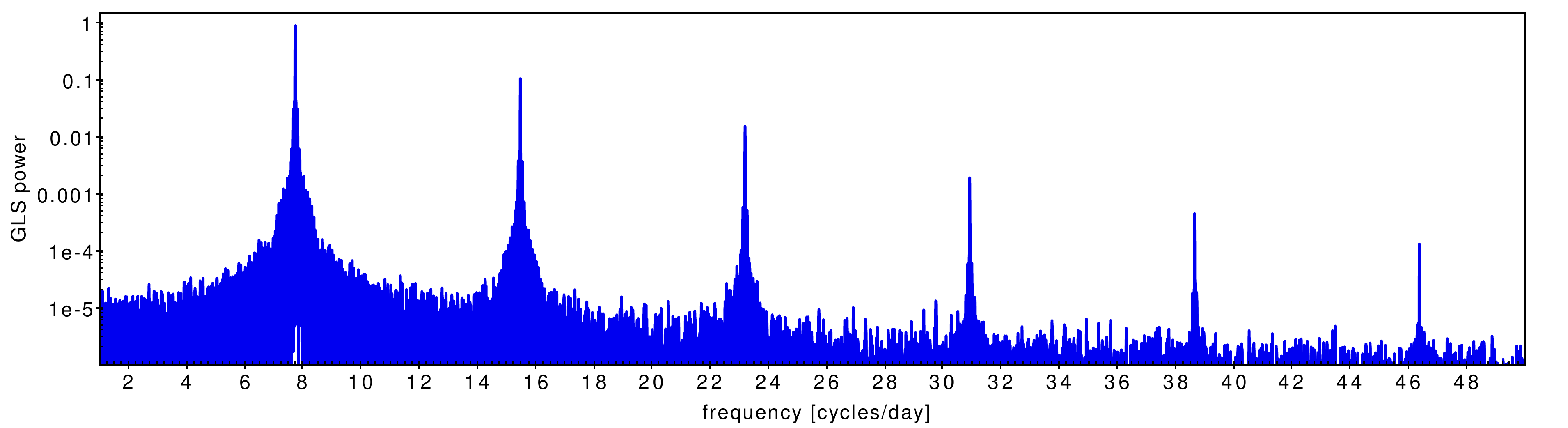}
    \caption{TESS photometry of Chang~134 = TYC~9158-919-1. \emph{Upper panel:} One-day section of the light curve from TESS sector 1. \emph{Lower panel:} GLS periodogram of the whole light curve including nine TESS sectors, stitched together and filtered as explained in Sect.~\ref{Section:Harmonic_analysis}.}
    \label{fig:lc_chang134}
\end{figure*}

\begin{figure*}
    \centering
    \includegraphics[width=0.95\textwidth]{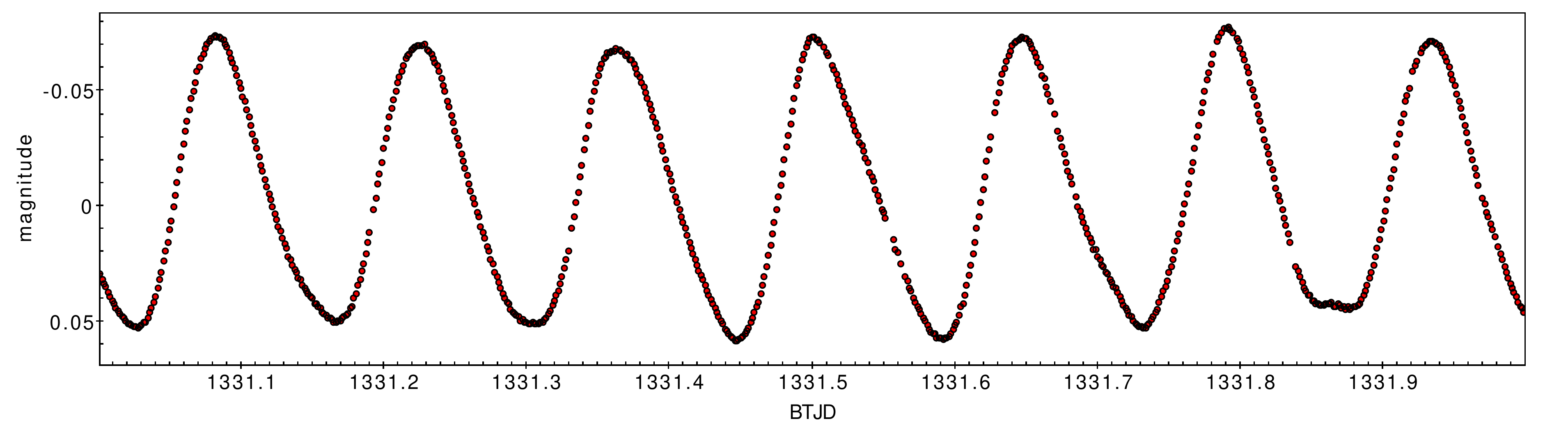}

    \includegraphics[width=0.95\textwidth]{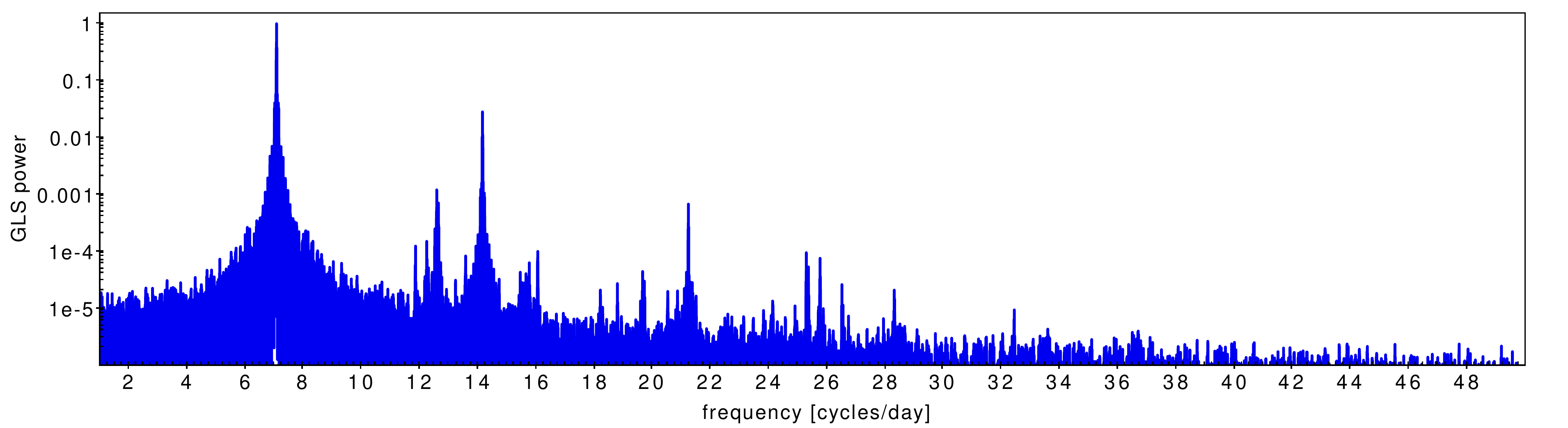}
    \caption{TESS photometry of V393~Carinae = HD 66260. \emph{Upper panel:} One-day section of the light curve from TESS sector 1. \emph{Lower panel:} GLS periodogram of the whole light curve including 14 TESS sectors, stitched together and filtered as explained in Sect.~\ref{Section:Harmonic_analysis}.}
    \label{fig:lc_v393}
\end{figure*}

\subsection{Frequency identification}
\label{Subsection:GLS}
In order to identify the most significant pulsation modes of our targets we made use of the Generalized Lomb-Scargle periodogram (GLS, \citealt{Zechmeister2018}), an improved version of the Lomb-Scargle periodogram (LS, \citealt{VanderPlas2018}) which is less sensitive to aliasing and returns more accurate frequencies. Because we searched for LTTE effects, all the frequencies might in principle be slightly varying as a function of time due to the phase shift itself. At this stage, we therefore need to calculate the average value over the full span of our observations for each frequency. This can be accomplished by stitching all the normalized orbits into a single light curve and running GLS on it. We limited our search from 1 to 50 cycles per day, thus including the typical frequency range of \ds{} modes \citep{Balona2011} and being well within the Nyquist limit for our sampling rate ($\sim$ 360~cycles per day).  The output are two periodograms for Chang~134 and V393~Car that are visually similar to those shown in Fig.~\ref{fig:lc_chang134} and \ref{fig:lc_v393}, but with a much higher S/N.
\valentina{In order to improve the estimate of the pulsation frequencies, the light curve is usually fitted by a multisinusoidal function using the periodogram peaks as a starting point (\citealt{Silvotti2018}). In our case, however, this additional step is not essential given the extremely high S/N of our data.}

\indent By identifying the strongest frequencies $\nu_i$ (and corresponding periods $P_i$) as the most prominent peaks in the periodogram, we found one dominant frequency for Chang 124 that was followed by its multiple integer harmonics, sorted hereafter by decreasing GLS power,
\begin{equation}\label{eq:freq_chang134}
\left\{ \begin{array}{cc}
P_0 = 0.12942447\;\textrm{d}\\ 
P_1 = 0.06471223\;\textrm{d}\\
P_2 = 0.04314149\;\textrm{d} 
\end{array}
\right .\quad
\left\{ \begin{array}{cc}
\nu_0 = 7.72651415\;\textrm{d}^{-1} \\ 
\nu_1 = 15.4530283\;\textrm{d}^{-1} \\
\nu_2 = 23.1795424\;\textrm{d}^{-1} 
\end{array}
\right . \textrm{ ,}
\end{equation}
and in the same way for V393~Car,
\begin{equation}\label{eq:freq_v393car}
\left\{ \begin{array}{cc}
P_0 = 0.14129519\;\textrm{d}\\ 
P_1 = 0.07064759\;\textrm{d}\\
P_2 = 0.04709840\;\textrm{d} 
\end{array}
\right .\quad
\left\{ \begin{array}{cc}
\nu_0 = 7.07738170\;\textrm{d}^{-1} \\ 
\nu_1 = 14.1547634\;\textrm{d}^{-1} \\
\nu_2 = 21.2321451\;\textrm{d}^{-1} 
\end{array}
\right . \textrm{ .}
\end{equation}
\valerio{A complete list of all the significant frequencies detected for Chang~134 and V393~Car can be found in Table~\ref{table:freq_chang134} and \ref{table:freq_v393car}, respectively.}

For both stars, we obtained $\nu_1 = 2\nu_0$ and $\nu_2 = 3\nu_0$, that is, the three most prominent frequencies are the second and third harmonics of the main pulsation mode. Chang~134 does not show any significant peak outside the harmonic series of $\nu_0$. The GLS power of the strongest peak for V393 Car outside the harmonics of $\nu_0$ (at $\sim$12.58288~d$^{-1}$) is three orders of magnitude smaller than the main mode. While irrelevant for our analysis due to its weakness, we note that this minor peak, detected at very high confidence in the TESS data, confirms the presence of an additional pulsation mode at 12.58 or 13.58 cycles per day claimed by \citet{Helt1984} that was never confirmed by subsequent works \citep{Garcia2001,Axelsen2014}. \valentina{The very clean window function of the TESS data enables us to state that 13.58 was just a one-day alias of the real peak at 12.58 cycles per day.} 


\subsection{Harmonic fit}

After the average frequencies $\nu_i = 2\pi/P_i$ of the most powerful pulsation modes were identified, we proceeded to fit a harmonic series in the form of Eq.~\ref{eq:harmser} to our light curves. As several TESS orbits are affected by long-term systematic errors even after the PDCSAP treatment (on a timescale of typically days to weeks), we added a polynomial term as a function of time to our model to mitigate their impact. After some trial and error, we concluded that a polynomial of order 2 (i.e., a quadratic baseline) is enough to provide us with an accurate fit. The full model $M$ we fit is therefore
\begin{equation}\label{eq:model}
M = b_0 + b_1(t-t_m) + b_2(t-t_m)^2 + \sum_{i=0}^{N-1} f_i(A_i,P_i,\phi_i)
,\end{equation}
where $t_m$ is the median time of each orbit, set to minimize the correlations between the fit $b_i$ parameters. The $P_i$ periods must be fixed to their average values determined in the previous section in order to obtain meaningful phase shifts; for $N$ harmonic components included, the number of free parameters is therefore $3+2N$. While $N$ might in principle be arbitrarily large, for instance, when all the components that are even barely detectable in the frequency spectrum are included, almost all the information on the phase shift is contained in the three most significant peaks, which sum up to $\gg 99\%$ of the total power. We therefore adopt $N=3$ for our model by including the $P_i$ constants from Eq.~\ref{eq:freq_chang134} and \ref{eq:freq_v393car}, implying nine free parameters for our fit. We verified this assumption by adding further harmonic terms as a test, which resulted in statistically indistinguishable $O-C$ diagrams at the expense of a much more intensive computation.

We fit the model in Eq.~\ref{eq:model} through a differential evolution Monte Carlo Markov chain algorithm (DE-MC; \citealt{TerBraak2006}), in which multiple MC chains are run in parallel and learn from each other how to converge to a global minimum in the parameter space, rather than running independently as in the classical approach. With respect to the least-squares techniques used by most previous works to fit harmonic series to light curves, DE-MC allowed us to obtain much more reliable error estimates on the final best-fit parameters because the posterior distribution takes any correlation between them into account. In order to increase the efficiency, we first fit only the oscillating term $\sum_if_i$ of our model to each orbit with a Levenberg-Marquardt method \citep{More1978}, to obtain reasonably good starting points for $A_i$ and $\phi_i$. Then we ran the DE-MC code, again orbit by orbit, to fit the full model with uninformative priors centered on the $A_i$, $\phi_i$ values found above, while the $b_i$ parameters were initially set to zero. After $1\,000\,000$ steps, the first 10\% of the chain was discarded as burn-in phase and the posterior distributions of the nine fit parameters were examined to ensure that convergence was reached.

At this stage, when we plotted the residuals of the light curves after the best-fit model had been subtracted, we realized that a non-negligible number of outliers had escaped the previous filtering steps. 
These data points are mostly located close to momentum dumps of the spacecraft or at the beginning of each orbit when  
TESS cameras are not yet thermally stabilized. To deal with them, we discarded all the outliers at more than 5 $\sigma$ and fed the clipped light curves for a second run of the DE-MC fit, with the same configuration. The final best-fit parameters, especially 
the phases $\phi_i$ to which we are most interested in, are statistically consistent with those delivered from the first run, but usually with slightly smaller uncertainties. 

Finally, the phase shifts $\phi _i$, by our definition within the $[0,1]$ interval, were converted into $O-C$ delays through a $P_i$ factor. \valerio{In principle, if the phase shifts were associated with independent pulsation modes and due solely to an LTTE, all the $\phi _i$ computed for the same orbit should result in the same $O-C$ value, within the measurement errors: $\phi_0P_0 = \phi_1P_1 = \phi_2P_2$, and computing a weighted average $\langle\phi_iP_i\rangle$ would be the rigorous way to obtain the overall phase shift. This would also be a great opportunity to independently verify whether the LTTE model is the best explanation because different modes should yield the same $O-C$  \citep{Hermes2018}. In our case, however, all the $\phi _i$ come from a single pulsation mode, and the harmonics do not carry additional information with respect to the fundamental frequency $\nu_0$, $P_0$, where most of the spectral power lies: $\sigma(\phi_0)\ll \sigma(\phi_1)\ll \sigma(\phi_2)$ and thus $\phi_0P_0\simeq \langle\phi_iP_i\rangle$.}  
For this reason, after confirming that the assumption above is true, we defined $\phi_0P_0$ as our $O-C$. We associated the median epoch of the orbit $t_m$ as a time stamp for each value as introduced above: $(O-C)(t_m) = \phi_0 \times P_0$.
The resulting $O-C$ diagrams for Chang~134 and V393~Car are plotted in Fig.~\ref{fig:oc_chang134} and Fig.~\ref{fig:oc_v393car}. The formal error bars are 1.5~s on average (range: 1.1-2.6~s) and 0.16~s (range: 0.14-0.19~s), respectively.


\section{LTTE modeling}
\label{Section:Orbital_solutions}

Even a quick look at both our $O-C$ diagrams (Fig.~\ref{fig:oc_chang134} and \ref{fig:oc_v393car}) reveals that neither can be reasonably fit by a straight line, that is, the resulting reduced $\chi^2_r$ is $\gg$1. The presence of a nonzero second derivative implies that the phase of the signal is evolving with time, and we can interpret this modulation by assuming that it is due to an LTTE from unseen companions. 
Following \cite{Kepler1991}, we parameterize our model as
\begin{equation}
    O-C = (t_{0} + \Delta P\cdot t) + \frac{1}{2}\frac{\dot{P}}{P}t^{2} + \alpha\cos \left(\frac{2\pi t}{P_{\text{orb}}} + \varphi \right) ,
    \label{eq:O-C_JJHermes} 
\end{equation}
where $t_{0}$ is an arbitrary reference time, $\Delta P$ is the difference between the \emph{\textup{actual}} average pulsation period $P$ and that estimated in Eq.~\ref{eq:freq_chang134} and \ref{eq:freq_v393car}, $\dot{P}=\mathrm{d}P/\mathrm{d}t$ accounts for a linear variation of $P$ as a function of time, and $\alpha$ is the amplitude of an LTTE signal  from a perturber on a circular\footnote{Throughout this paper we assume a circular orbit as a simplifying hypothesis for our LTTE model, although a significant fraction of exoplanetary systems and binary stars are known to have nonzero eccentricity \citep{Kim2018}. This is justified by our need to avoid two additional free parameters ($e$ and $\omega$) and is further discussed in Section~\ref{Section:Discussion_and_conclusions}.} 
orbit having an orbital period $P_\mathrm{orb}$ and with a phase $\varphi$. 

From a physical point of view, the first term of Eq.~\ref{eq:O-C_JJHermes} does not carry information because it represents just a change in the slope of the $O-C$ and does not imply any variation of the pulsation period. The second term, \valerio{in the LTTE framework,} represents a constant acceleration with respect to the barycenter of the system, possibly caused by a massive, perturbing body whose $P_\mathrm{orb}$ is much longer than our observing baseline ($\approx$1000~d), such as stellar companions on very wide orbits. \valerio{We should be aware here that several physical mechanisms other than LTTE can result in a $\dot P \ne 0$ term, including nonlinear interactions between different pulsation modes and stellar evolution effects. We return on this point in Section~\ref{sec:sol_v393char}.}

\valerio{Last, the third, oscillating term of Eq.~\ref{eq:O-C_JJHermes}} is the LTTE modulation we are most interested in. As the first two terms play the role of a quadratic baseline in our fit, we make it more explicit by rewriting the model with $a_0=t_0$, $a_1=\Delta P$ and $a_2=0.5(\dot P/P)$,
\begin{equation}\label{eq:model_quad}
    (O-C)_\textrm{quad} = a_0 + a_1 (t-t_m) +a_2 (t-t_m)^2  + \alpha\cos \left(\frac{2\pi t}{P_{\text{orb}}} + \varphi \right) 
,\end{equation}
where the median epoch of each orbit $t_m$ has been subtracted from the time variable in order to minimize the correlations between $a_0$, $a_1$, and $a_2$. 

In order to confirm the uniqueness of our best-fit solution and the significance of the quadratic coefficient $a_2$, we also performed a second fit to each data set by fixing $a_2=0$ in our model, that is, by imposing a linear baseline,
\begin{equation}\label{eq:model_lin}
    (O-C)_\textrm{lin} = a_0 + a_1 (t-t_m) + \alpha\cos \left(\frac{2\pi t}{P_{\text{orb}}} + \varphi \right) .
\end{equation}

According to the \citet{Irwin1952} model, for a circular orbit (eccentricity $e=0$) and assuming that the perturbing body is much less massive than the star ($m_\mathrm{p}\ll M_\star$), the amplitude of the LTTE signal is 
\begin{equation}
\alpha = \frac{a \sin(i)}{c}\,\frac{m_{\text{p}}}{ M_{\star}} \textrm{ ,} \label{eq: semi_amplitude_LTTE} 
\end{equation}
where \textit{c} is the speed of light, \textit{a} is the orbital semimajor axis of the perturber and \textit{i} is the inclination of its orbital plane with respect to the sky plane. In other words, when $\alpha$ and $P_\mathrm{orb}$ are measured from the $O-C$ data, the stellar mass is known (from Table~\ref{tab:stellar_parameters}) and $a$ is derived from Kepler's laws, we can constrain the minimum mass of the perturber as $m_\mathrm{p}\sin i$, with the same degeneracy on $i$ as for the RV and astrometric techniques,
\valerio{
\begin{equation}\label{msini}
    m\sin(i) = \alpha \times c\left ( \frac{M_\star}{P_\mathrm{orb}} \right )^{2/3} \left ( \frac{G}{4\pi^2}\right ) ^{-1/3}\textrm{ .}
\end{equation}}

We started the fitting process by computing a GLS periodogram on our $O-C$ diagrams to search for periodic modulations. For both targets, a single, well-defined peak stands out, at about 82~d and 1000~d for Chang~134 and V393~Car, respectively. We then fit our full LTTE model with both a quadratic and linear baseline (parameterized as in Eq.~\ref{eq:model_quad} and \ref{eq:model_lin}, respectively) on our $O-C$ diagrams through DE-MC, setting uninformative priors on all the five or six free parameters involved: $a_0$, $a_1$, $a_2$, $\alpha$, $P_\mathrm{orb}$, and $\varphi$ in the $(O-C)_\textrm{quad}$ model, and $a_0$, $a_1$, $\alpha$, $P_\mathrm{orb}$, and $\varphi$ in the $(O-C)_\textrm{lin}$ model, and centering the boundaries of $P_\mathrm{orb}$ on the values previously found by GLS to speed up the convergence. As done at the harmonic analysis stage, after $1\,000\,000$ steps, the first 10\% of the chain was then discarded as burn-in phase, and we extracted the best-fit values and the associated errors of our parameters from their posterior distributions. All these values are reported in Table~\ref{tab:LTTE_parameters} for the four independent fits. The mass of the perturbing body $m_\mathrm{p}\sin i$ was calculated as well as a derived parameter by propagating the relevant errors in Eq.~\ref{msini}.

\begin{table*}\centering
      \caption[]{Output parameters and associated errors for the best-fit LTTE models (Eq.~\ref{eq:O-C_JJHermes}) found for Chang~134 and V393~Car. \valentina{For each of the four independent fits (i.e., linear and quadratic), we report the best-fit values and uncertainties of our parameters returned by their posterior distributions as a result of the DE-MC iterations.}}
         \begin{tabular}{l|rcl|rcl|rcl|rcl}
            \multicolumn{1}{c}{} & \multicolumn{3}{c}{\textbf{Chang 134}} & \multicolumn{3}{c}{\textbf{Chang 134}} & \multicolumn{3}{c}{\textbf{V393 Car}}&
            \multicolumn{3}{c}{\textbf{V393 Car}}\\
            \hline\hline
            baseline & \multicolumn{3}{|c|}{linear} & \multicolumn{3}{|c|}{quadratic} & \multicolumn{3}{|c|}{linear}& \multicolumn{3}{|c}{quadratic}\\
            $t_m$ $[\textrm{BJD}_\textrm{TDB}]$ & \multicolumn{3}{|c|}{1668.16969} & \multicolumn{3}{|c|}{1668.16969} & \multicolumn{3}{|c|}{2060.26634}&
            \multicolumn{3}{|c}{2060.26634}\\
            $a_{0}$ [s] & 0.17 & $\pm$ & 0.35 & 0.19  & $\pm$ & 0.99 & $-$14.63 & $\pm$ & 0.19 & $-$68.22 & $\pm$ & 0.26 \rule{0pt}{15pt}\\
            $a_{1}$ [ppt] & 1.4 & $\pm$ & 1.2 &  1.9 & $\pm$ & 1.8 & 205.25 & $\pm$ & 0.23 & 448.3 & $\pm$ & 0.14 \\
            $a_{2}$ [ppm/s] & \multicolumn{3}{c|}{0 (fixed)} & $-2.2$ & $\pm$ & 7.5 & \multicolumn{3}{c|}{0 (fixed)} & 650.3 & $\pm$ & 3.7 \\
            $P_{\text{orb}}$ [days] & 81.96 & $\pm$ & 0.28  & 82.09 & $\pm$ & 0.29 & 1071.5 & $\pm$ & 1.0 & 724.7 & $\pm$ & 2.0 \rule{0pt}{15pt}\\
            $\phi_{0}$ & 0.204 & $\pm$ & 0.072 & 0.215 & $\pm$ & 0.074 & 0.7485 & $\pm$ & 0.0016 & 0.9081 & $\pm$ & 0.0071 \\
            $\alpha$ [s]& 6.11 & $\pm$ & 0.65 & 6.99 & $\pm$ & 0.66 & 107.25 & $\pm$ & 0.09 & 47.96 & $\pm$ & 0.17 \\
            $m_\mathrm{p}\sin i$ [$M_\textrm{jup}$]& 43.1 & $\pm$ & 4.7 & 49.2 & $\pm$ & 4.8 & 175.7 & $\pm$ & 9.2 & 102.0 & $\pm$ & 5.3 \rule{0pt}{15pt}\\
            $\chi^2$ & \multicolumn{3}{c|}{56.7} & \multicolumn{3}{c|}{56.4} & \multicolumn{3}{c|}{17326} & \multicolumn{3}{c}{8627}  \\
            $\chi^2_\mathrm{r}$ & \multicolumn{3}{c|}{4.36} & \multicolumn{3}{c|}{4.69} &\multicolumn{3}{c|}{753} & \multicolumn{3}{c}{392}  \\
 \hline
         \end{tabular}\label{tab:LTTE_parameters}
\tablefoot{As discussed in Section~\ref{Section:Orbital_solutions}, the adopted solution for Chang~134 is based on the linear model (first column), while for V393~Car, the data coverage is still insufficient to tell which model between $(O-C)_\textrm{lin}$ and $(O-C)_\textrm{quad}$ corresponds to the actual orbital solution.}
   \end{table*}

\subsection{Orbital solution for Chang~134}\label{sec:sol_chang134}

\begin{figure*}
    \centering
    \includegraphics[width=0.75\textwidth, trim={0 20 0 0}, clip]{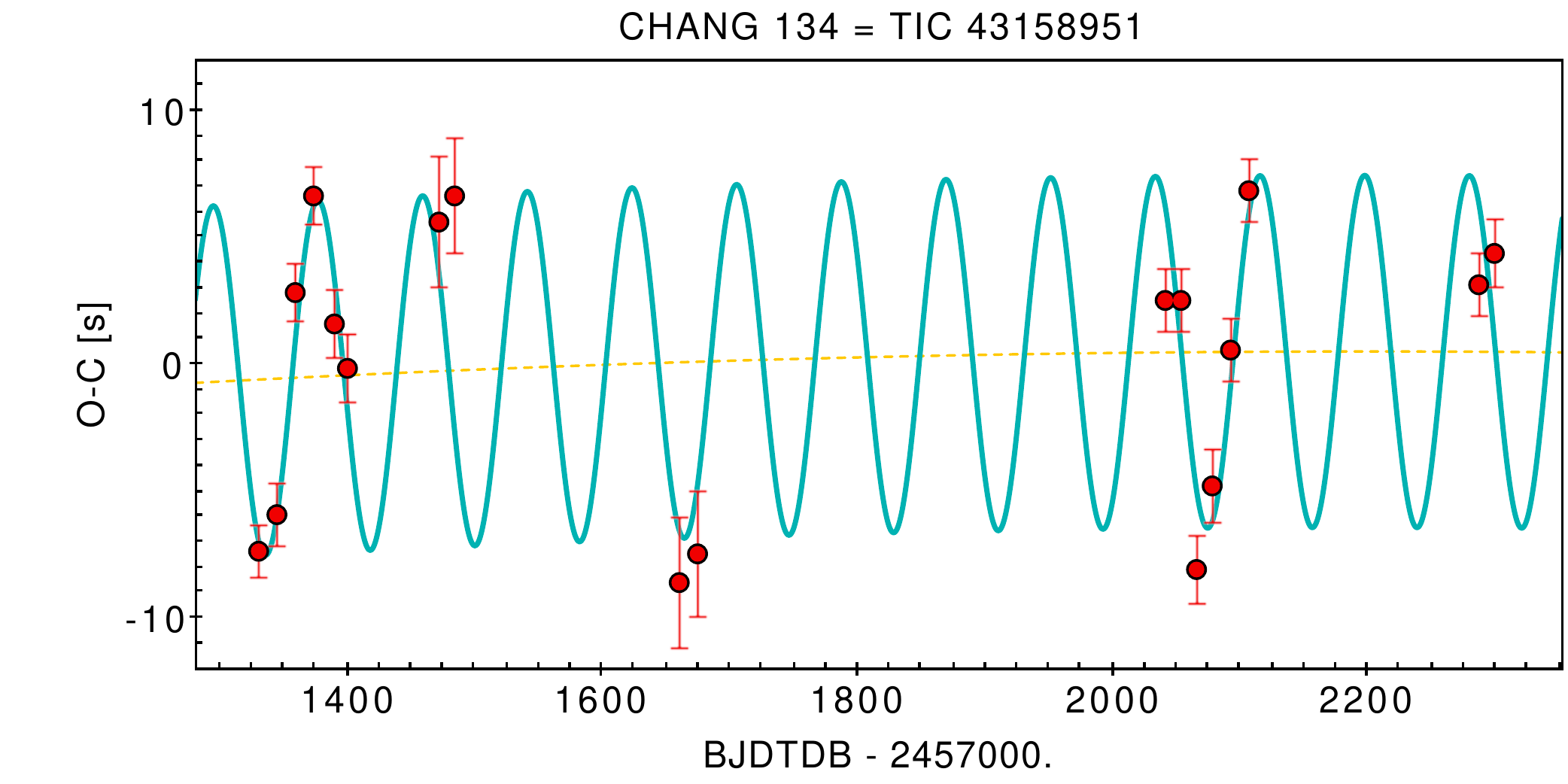}

    \includegraphics[width=0.75\textwidth, trim={0 0 0 0}, clip]{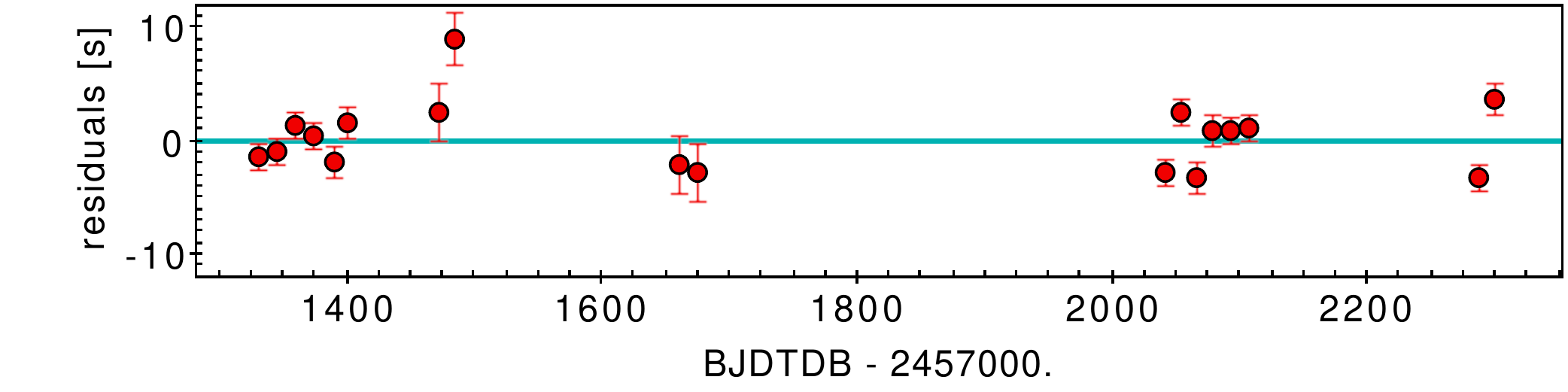}
    \caption{Orbital solution for Chang~134. \emph{Upper panel:} O-C diagram for the main pulsation mode of Chang~134 in seconds. The best-fit LTTE solution (including a quadratic trend) is overplotted as a continuous cyan line, and the quadratic baseline is plotted as a dashed orange line (see Section~\ref{Section:Harmonic_analysis} for details). \emph{Lower panel:} Residuals from the best-fit model in seconds.}
    \label{fig:oc_chang134}
\end{figure*}

The best-fit $(O-C)_\textrm{quad}$ model for Chang~134 is overplotted on our $O-C$ in the upper panel of Fig.~\ref{fig:oc_chang134} as a solid cyan line, with the quadratic baseline $a_0 + a_1 t +a_2 t^2$ plotted as a dashed orange line. The resulting $\chi^2$ of the residuals (lower panel of Fig.~\ref{fig:oc_chang134}) is 56.7, with the reduced $\chi^2$ being $\chi_{\text{red}}^2 = \chi^2/\textrm{DOF} \simeq 4.36$, where $\textrm{DOF} = 18 - 6 = 12 $ are the degrees of freedom of our fit. It is worth noting that the best-fit value for $a_2$ is statistically consistent with zero (Table~\ref{tab:LTTE_parameters}, second column), that is, we do not see any evidence of long-term LTTE effects because the derived $\dot P$ is also consistent with zero. This is confirmed by the fact that the best-fit linear model $(O-C)_\textrm{lin}$ (Table~\ref{tab:LTTE_parameters}, first column) yielded the same results within the error bars for all the fit parameters, with a virtually indistinguishable $\chi^2$ (56.4 vs.~56.7). In other words, because the $(O-C)_\textrm{quad}$ model is not favored by our data, 
we hereafter adopt the results from the $(O-C)_\textrm{lin}$ fit.

The best-fit returned the amplitude of the LTTE term to be $\alpha = 6.11 \pm 0.65$~s (according to Eq.~\ref{eq:O-C_JJHermes}), and after propagating the errors through Eq.~\ref{eq: semi_amplitude_LTTE}, we obtain a minimum mass of $m_\mathrm{p}\sin i=43.1\pm 4.7$~$M_\textrm{jup}$ for the hypothetical companion. This is within the brown dwarf regime (13-80~$M_\textrm{jup}$; \citealt{Grieves2021}).

\subsection{Orbital solution for V393~Car}\label{sec:sol_v393char}

\begin{figure*}
    \centering
    \includegraphics[width=0.75\textwidth, trim={0 20 0 0}, clip]{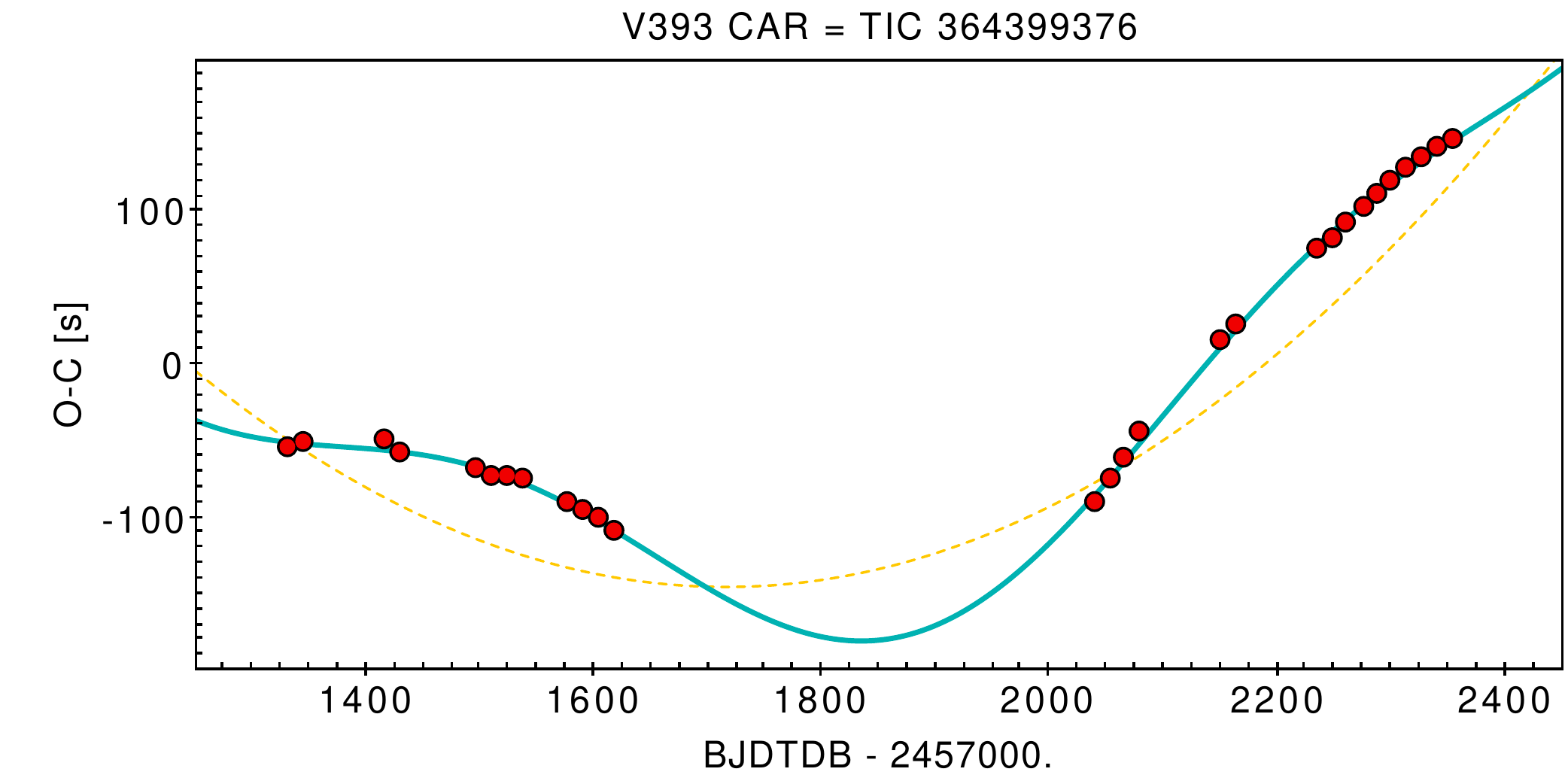}

    \includegraphics[width=0.75\textwidth, trim={0 0 0 0}, clip]{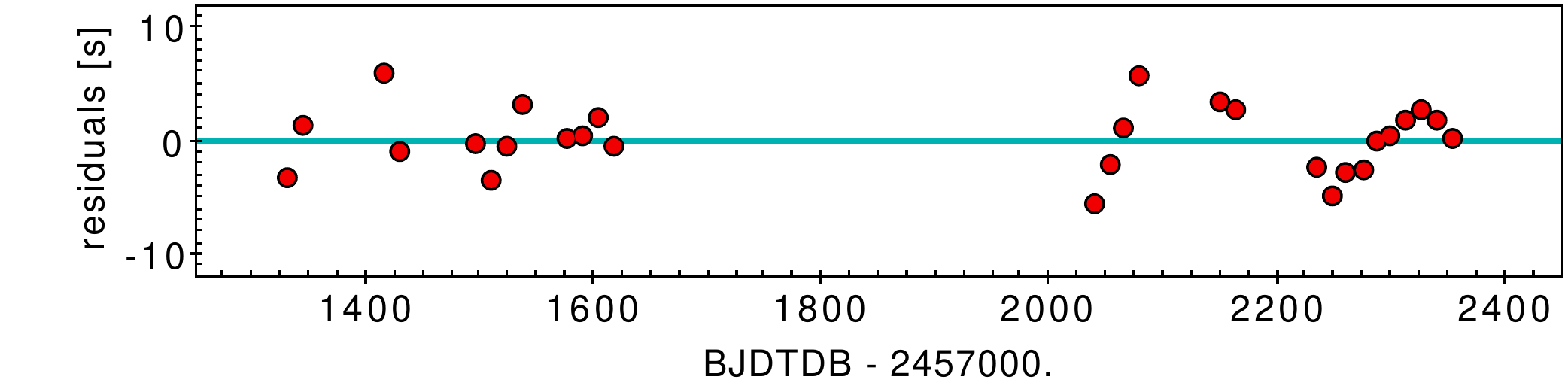}
    \caption{Quadratic orbital solution for V393~Car. \emph{Upper panel:} O-C diagram for the main pulsation mode of V393~Car in seconds. The best-fit LTTE solution with a quadratic baseline $(O-C)_\textrm{quad}$ (Eq.~\ref{eq:model_quad}) is overplotted as a continuous cyan line (see Section~\ref{Section:Harmonic_analysis} for details). The quadratic baseline $a_0 + a_1 t +a_2 t^2$ is plotted separately as a dashed orange line. The nominal error bars ($\sim$0.16~s on average) are smaller than the point size. \emph{Lower panel:} Residuals from the best-fit model in seconds.}
    \label{fig:oc_v393car} 
\end{figure*}

\begin{figure*}
    \centering
    \includegraphics[width=0.75\textwidth, trim={0 20 0 0}, clip]{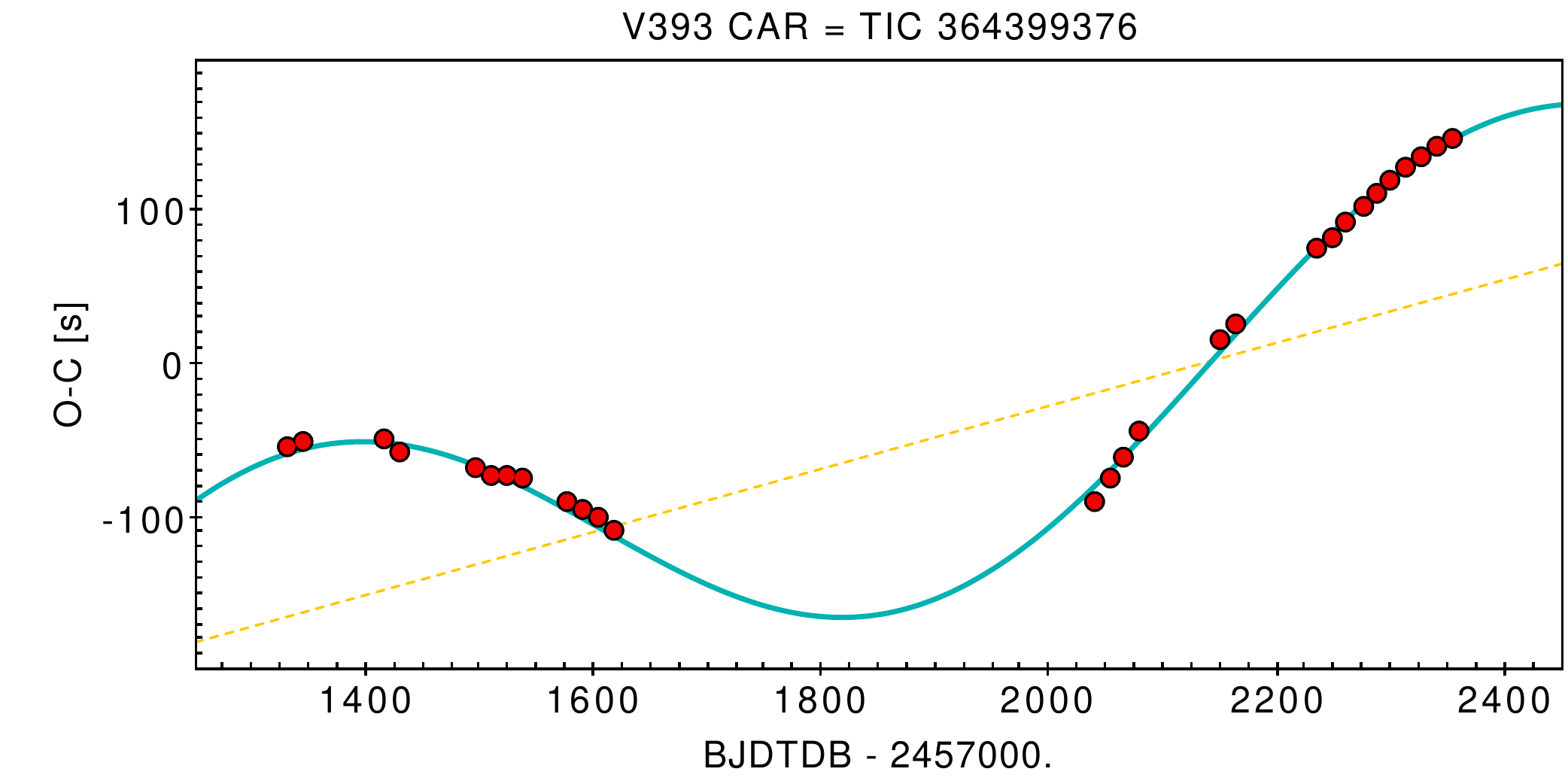}

    \includegraphics[width=0.75\textwidth, trim={0 0 0 0}, clip]{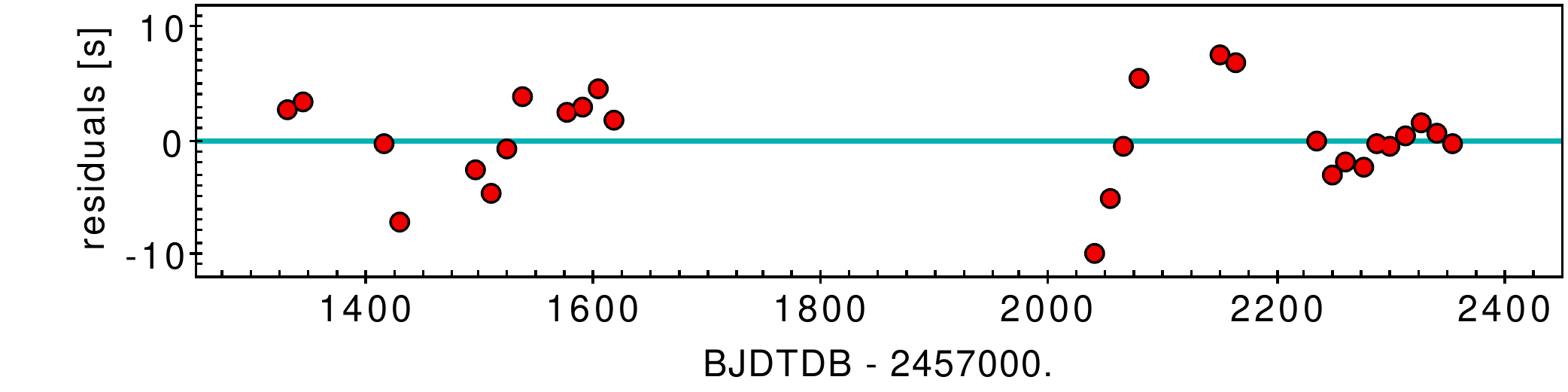}
    \caption{Linear orbital solution for V393~Car. \emph{Upper panel:} Same as Fig.~\ref{fig:oc_v393car}, but with the best-fit LTTE solution with a linear baseline $(O-C)_\textrm{lin}$ (Eq.~\ref{eq:model_lin}). The linear baseline $a_0 + a_1 t$ is plotted separately as a dashed orange line. \emph{Lower panel:} Residuals from the best-fit model in seconds.}
    \label{fig:oc_v393car_lin}
\end{figure*}

Unlike the case of Chang~134, 
the results from 
V393~Car fits are more difficult to interpret 
because the two models $(O-C)_\textrm{quad}$ and $(O-C)_\textrm{lin}$, while having a similar shape, led us to different orbital solutions. The two best-fit models are overplotted on our $O-C$ in the upper panel of Fig.~\ref{fig:oc_v393car} and \ref{fig:oc_v393car_lin} as a solid cyan line, respectively, with the linear and quadratic baseline $a_0 + a_1 t +a_2 t^2$ plotted as a dashed orange line. 

While it is clear that at least an oscillating term is needed to fit the $O-C$ data, the inclusion of a quadratic term yields different parameters for the LTTE sinusoid: $P\simeq 1071$~d with amplitude $\alpha\simeq 107$~s for the  $(O-C)_\textrm{lin}$ fit versus~ $P\simeq 723$~d with amplitude $\alpha\simeq 48$~s for the $(O-C)_\textrm{quad}$ fit. The discrepancy of a factor of $\sim$2 in the $\alpha$ parameter translates into a similar ratio for the derived hypothetical companion masses: $m_\mathrm{p}\sin i=175.7 \pm 9.2$~$M_\textrm{jup}$, or $0.17 \pm 0.01$~$M_\odot$ (linear model) and $m_\mathrm{p}\sin i=102.0 \pm 5.3$~$M_\textrm{jup}$, or $0.10 \pm 0.01$~$M_\odot$ (quadratic model). We emphasize that for both scenarios, the minimum mass would be consistent with a cool dwarf of spectral type M5V and M6V, respectively \citep{Pecaut2013}.

A technique that is commonly employed for model selection problems like this is the evaluation of metrics such as the Akaike information criterion (AIC; \citealt{Akaike1974}) or the Bayesian information criterion (BIC; \citealt{Schwarz1978}). In the case of Gaussian distributions, the latter can be written as $\textrm{BIC}=\chi ^2 + k\ln n$, where $k$ is the number of free parameters and $n$ is the number of data points. If applied literally to our problem, the BIC would overwhelmingly favor our $(O-C)_\textrm{quad}$ scenario ($\Delta\textrm{BIC}\gg 10$). Unfortunately, this result is distorted by the fact that the $\chi^2$ of the residuals for both fits (lower panel of Fig.~\ref{fig:oc_v393car} and \ref{fig:oc_v393car_lin}) is exceptionally high due to the extremely low formal error bars with respect to the systematic errors at play (on which we comment in Section~\ref{Section:Discussion_and_conclusions}): for the $(O-C)_\textrm{quad}$ fit, for instance, we obtained $\chi^2=8627$, with the reduced $\chi^2$ being $\chi_{\text{red}}^2 = \chi^2/\textrm{DOF} \simeq 392$, where $\textrm{DOF} = 28 - 6 = 22 $ are the degrees of freedom of our fit. This fact, and the very sparse phase sampling of our signal, prevents us from reaching a firm conclusion on the properties of this candidate companion, which will require further data to be confirmed and better constrained. New TESS observations of V393~Car are planned for Cycle 5, during four consecutive sectors (61-64; from January to May 2023), and again in Sector 68 (August 2023). Since the predictions from the $(O-C)_\textrm{quad}$ and $(O-C)_\textrm{lin}$ models will diverge by $\sim$12~h by mid-2023, we expect that the extension of our analysis to the new TESS light curves will definitely be conclusive about the correct scenario.

\valerio{As a side note, we could wonder whether the quadratic term we found when we assumed a quadratic baseline ($a_2=650\pm 4$~ppm/s as fitted on our $O-C$ plane, translating into $\dot P/P\simeq 5.5\cdot10^{-6}$~$\textrm{yr}^{-1}$) might be consistent with other non-LTTE mechanisms, as previously mentioned. Significant nonlinear interactions between pulsating modes \citep{Silvotti2018,Bowman2021} are not at play here because the spectral power outside the main pulsation mode and its harmonics is essentially negligible. On the other hand, the typical $\dot P/P$ expected from evolutionary effects range from $10^{-9}$ to $10^{-7}$~$\textrm{yr}^{-1}$ \citep{Breger1998,Xue2022}, that is, it is much lower than our fitted value. }

\valerio{The same reasoning can be applied to Chang~134, for which we measure a nonsignificant quadratic term $a_2=-2.2\pm 7.5$~ppm/s corresponding to $\dot P/P\simeq (-1.8\pm 6.6)\cdot10^{-8}$~$\textrm{yr}^{-1}$. Unlike V393~Car, this upper limit is still within the predictions for evolutionary effects. For instance, if we adopt $\dot P/P\simeq 3\cdot10^{-9}$~$\textrm{yr}^{-1}$ (as measured on AE~UMa by \citealt{Xue2022}), we obtain a total contribution on the $O-C$ of $\sim0.35$~s over the full baseline of our observations ($\sim1000$~d), which is well below our current measurement error.}

\section{Discussion and conclusions}
\label{Section:Discussion_and_conclusions}

We applied the PT technique to two \ds{} stars observed by TESS in a large number of sectors, detecting a periodic modulation on their $O-C$ that is consistent with the presence of companions in both cases: a BD ($m_\mathrm{p}\sin i=43.1\pm 4.7$~$M_\textrm{jup}$) on a $P\simeq 82$~d orbit around Chang~134, and a more massive body ($m_\mathrm{p}\sin i\simeq0.10$-0.17~$M_\odot$) around V393~Car, whose orbital parameters are still not firmly constrained by the limited phase sampling of the signal. This double detection on the very first two targets analyzed by our project may appear to be very lucky. Still, the
binary fraction of \ds{} primaries as estimated by a previous PT search on Kepler data is rather high: $15.4\pm 1.4\%$ \citep{Murphy2018}. In addition, unlike the quoted work, we initially restricted our sample in order to target only the most favorable stars in terms of sensitivity to LTTE signals. The overall binary fraction of A-type stars is estimated to be significantly larger than $50\%$ by most authors \citep{Duchene2013}, making our results less surprising. Any further statistical implication is prevented by the small size of our sample and by the heavy selection effects at play. As a side note, however, we mention that \citet{Borgnet2019} constrained the BD frequency within 2-3~au around AF stars to be below 4\% (1$\sigma$)  based on 225 targets
observed with SOPHIE and/or HARPS, and that other authors hypothesized the presence of a brown dwarf desert extending to early-type stars \citep{Murphy2018}. This would make the companion of Chang~134 an uncommon object, worthwhile to be followed up with more data and/or different techniques. The future TESS observations (at least five sectors of Cycle 5 are already scheduled, starting in March 2023) will also possibly help in constraining the eccentricity of the LTTE orbit, forced to zero in our analysis to avoid overfitting given our limited phase coverage and small number of $O-C$ points.

In addition to our results for these two specific stars, our analysis was based on an independent and improved implementation of the PT method \citep{Murphy2014}. It is also intended as a pilot study that is a prelude to a systematic search of LTTE companions around pulsating stars on a much larger scale, exploiting the huge sample of \ds{} targets that are and will be monitored through the availability of TESS FFIs at 10-minute cadence. The results for Chang~134, in particular, demonstrates that the detection of bodies in the substellar regime is perfectly feasible from 
TESS light curves, and that in some particularly favorable configuration, candidates with planetary masses ($m_\mathrm{p}\sin i\lesssim 13$~$M_\textrm{jup}$) might be already within reach. A perturber with the same LTTE amplitude as was found on Chang~134 ($\alpha\simeq 7$~s), for instance, but on a larger orbit at $P\simeq 800$~d, would imply a companion with a mass of $\simeq 2$~$M_\textrm{jup}$ detected at high significance.

On the other hand, our analysis highlighted several limiting factors that are worth discussing further. The first factor, largely self-evident in the case of V393~Car (Section~\ref{sec:sol_v393char}), is related 
to the \emph{\textup{sparse sampling}} of the TESS light curves, an unavoidable consequence of its fixed scanning law \citep{Ricker2015}. Even for targets within or close to the CVZ at ecliptic latitudes $|\beta|\gtrsim 78^\circ$, that is, targets that are observed in up to 13 contiguous sectors, there is always a large gap every other year, which make the detection of signals at longer periods ($P\gtrsim 1$~year) difficult or ambiguous. Unfortunately, this is also the region in the parameter space where the PT technique is more sensitive to low-mass perturbers, owing to the Eq.~\ref{eq: semi_amplitude_LTTE} relation. As TESS is continuing its mission through extension phases,  
the accumulated data will mitigate this issue by filling out all or most of the orbital phases on the folded $O-C$ diagram.

A second, subtler limiting factor is the excess scatter in our $O-C$ data, as can be seen from the very high $\chi^2_\mathrm{r}$ values reported in Table~\ref{tab:LTTE_parameters} and discussed Section~\ref{sec:sol_v393char}. Our residuals from the best-fit LTTE model have an rms of $\sim 3$~s in both fits, regardless of the average nominal error bars, which are about ten times lower on V393~Car than on Chang~134. The most plausible cause for this are systematic errors in the\emph{\textup{ absolute calibration of the time stamp}}. This issue has been anticipated by the TESS team, and in particular by the TESS Asteroseismic Science Consortium (TASC), who gave accurate timing requirements in the  SAC/TESS/0002/6 document\footnote{\url{https://tasoc.dk/docs/SAC_TESS_0002_6.pdf}}, including RS-TASC-05: ``[\dots] the time given for each exposure  should be accurate over a period of 10 days to better than 1 second''. No requirement is given at longer timescales, however, and the TESS data release notes (DRN\footnote{\url{https://archive.stsci.edu/tess/tess_drn.html}}) have reported systematic offsets in the time stamps for several sectors on the order of seconds, that is, consistent with the excess $O-C$ scatter we measured. While some of the reported offsets have been already corrected in subsequent update data releases, others are still waiting to be implemented. A ground-based campaign by \citet{VonEssen2020} tried to independently confirm the absolute time calibration of TESS by comparing observations of a sample of eclipsing binaries in common, measuring a global offset of  $5.8\pm 2.5$~s; the errors of the measurements at individual epochs, however, are too large ($\gtrsim 10$~s) to confirm or disprove  the systematic errors we see. We emphasize that offsets of a few seconds are usually completely negligible when investigating transiting exoplanets, but they are crucial in extending the sensitivity of the PT technique to planet-mass companions. An important outcome of a large-scale PT analysis of TESS data will allow us also to identify and correct these systematic errors by comparing the timing residuals of tens or hundreds of targets, following a self-calibration approach. 

The same technique as applied in this work, and in general, the expertise gained through this project, will also help the scientific exploitation of PLATO \citep{Rauer2014}, for which the PT analysis could be a compelling case of ancillary science. Unlike TESS, no FFIs will be downloaded from PLATO during the nominal observing phase,  and no stars earlier than F5 are to be included in the main target samples, which are focused on transit search around solar-type stars \citep{Montalto2021}. Nevertheless, about $8\%$ of the PLATO science data rate will be allocated to the General Observer (GO) program through ESA calls open to the whole astrophysical community. Any \ds{} star (and, more in general, pulsating star) allocated as GO within a long-pointing field \citep{Nascimbeni2022} would result in a mostly uninterrupted light curve on a baseline of two to four years, with a cadence\footnote{The nominal cadence of the ``normal'' PLATO cameras will be 25~s, i.e., a factor of about 5 shorter than the TESS standard cadence \citep{Rauer2014}.} and photometric precision far better than what is achieved by TESS, overcoming the sparse phase sampling affecting the latter. Even more important, the time stamps of each PLATO data point will be accurate within 1~s in an absolute sense by formal scientific requirement because of the specifically designed calibration processes that also include the monitoring of a preselected set of detached eclipsing binaries \citep{Nascimbeni2022}.

\begin{acknowledgements}
The authors wish to thank the referee, Dr.~Roberto Silvotti, for the valuable and fruitful comments and suggestions which significantly improved our manuscript.
This research has made use of the SIMBAD database (operated at CDS, Strasbourg, France;\citealt{Wenger2000}), the VARTOOLS Light Curve Analysis Program (version 1.39 released October 30, 2020, \citealt{Hartman_and_Bakos_2016}), TOPCAT and STILTS \citep{Taylor2005,Taylor2006}.
Valerio Nascimbeni and Giampaolo Piotto recognize support by ASI under program PLATOASI/INAF
agreements 2015-019-R.1-2018.
\end{acknowledgements}

\noindent\emph{Note added in proof.} After the acceptance of this paper, we have been contacted by Dr.~D.~Hey who let us know that he found a similar orbital solution for V393~Car using the \texttt{maelstrom} code \citep{Hey2020}. In particular, his results suggest that the linear-baseline model yields a better fit the timing data.

\bibliographystyle{aa}
\bibliography{biblio}

\appendix
\section{O-C data tables}
\begin{table*}\centering
\caption{$O-C$ data points, best-fit models, and residuals for Chang 134.}
\begin{tabular}{crrrrrrr}
\hline
  \multicolumn{1}{c}{$\textrm{BJD}_\textrm{TDB}$}  &
  \multicolumn{1}{c}{$O-C$} &
  \multicolumn{1}{c}{$\sigma(O-C)$} &
  \multicolumn{1}{c}{$(O-C)_\textrm{lin}$} &
  \multicolumn{1}{c}{$(O-C)_\textrm{lin}$} &
  \multicolumn{1}{c}{$(O-C)_\textrm{quad}$} &
  \multicolumn{1}{c}{$(O-C)_\textrm{quad}$} \\
  \multicolumn{1}{c}{$-2457000.$}  &
  \multicolumn{1}{c}{[s]} &
  \multicolumn{1}{c}{[s]} &
  \multicolumn{1}{c}{model} &
  \multicolumn{1}{c}{residuals} &
  \multicolumn{1}{c}{model} &
  \multicolumn{1}{c}{residuals} \\
\hline\hline
  1331.92359 & $-$7.438& 1.068 & $-$6.108& $-$1.330& $-$6.022& $-$1.415 \\
  1345.90832 & $-$5.963& 1.198 & $-$5.058& $-$0.904& $-$4.998& $-$0.964 \\
  1360.60338 &  2.809& 1.122 &  1.510&  1.298&  1.574&  1.235 \\
  1375.00044 &  6.625& 1.101 &  6.235&  0.389&  6.288&  0.336 \\
  1390.56269 &  1.500& 1.381 &  3.572& $-$2.072&  3.571& $-$2.071 \\
  1401.41733 & $-$0.206& 1.291 & $-$1.552&  1.346& $-$1.603&  1.396 \\
  1473.17464 &  5.535& 2.578 &  3.383&  2.152&  3.223&  2.312 \\
  1484.59180 &  6.586& 2.316 & $-$2.045&  8.631& $-$2.253&  8.839 \\
  1660.82652 & $-$8.612& 2.526 & $-$6.003& $-$2.608& $-$6.403& $-$2.208 \\
  1675.51285 & $-$7.523& 2.504 & $-$4.268& $-$3.255& $-$4.614& $-$2.909 \\
  2042.19462 &  2.504& 1.243 &  5.634& $-$3.129&  5.442& $-$2.937 \\
  2054.87251 &  2.432& 1.233 &  0.210&  2.222& $-$0.032&  2.465 \\
  2065.98714 & $-$8.098& 1.382 & $-$4.555& $-$3.543& $-$4.765& $-$3.333 \\
  2079.35797 & $-$4.880& 1.406 & $-$5.615&  0.735& $-$5.701&  0.820 \\
  2093.48079 &  0.471& 1.232 & $-$0.462&  0.934& $-$0.423&  0.894 \\
  2107.49034 &  6.839& 1.251 &  5.656&  1.182&  5.709&  1.129 \\
  2287.53230 &  3.071& 1.236 &  6.014& $-$2.943&  6.356& $-$3.285 \\
  2300.70057 &  4.346& 1.289 &  0.490&  3.856&  0.779&  3.566 \\
\hline\end{tabular}\label{table:oc_chang134}
\tablefoot{The columns list the time stamp in $\textrm{BJD}_\textrm{TDB}$ \citep{Eastman2010} minus a constant, the $O-C$ value and its error $\sigma(O-C)$ as evaluated by the pipeline described in Section~\ref{Section:Harmonic_analysis}, the best-fit model with a linear baseline $(O-C)_\textrm{lin}$ (Eq.~\ref{eq:model_lin}) and its residuals, and the best-fit model with a quadratic baseline $(O-C)_\textrm{quad}$ (Eq.~\ref{eq:model_quad}) and its residuals. The unit of all columns except for the first one is seconds.}
\end{table*}

\begin{table*}\centering
\caption{$O-C$ data points, best-fit models, and residuals for V393~Car.}
\begin{tabular}{crrrrrr}
\hline
  \multicolumn{1}{c}{$\textrm{BJD}_\textrm{TDB}$}  &
  \multicolumn{1}{c}{$O-C$} &
  \multicolumn{1}{c}{$\sigma(O-C)$} &
  \multicolumn{1}{c}{$(O-C)_\textrm{lin}$} &
  \multicolumn{1}{c}{$(O-C)_\textrm{lin}$} &
  \multicolumn{1}{c}{$(O-C)_\textrm{quad}$} &
  \multicolumn{1}{c}{$(O-C)_\textrm{quad}$} \\
  \multicolumn{1}{c}{$-2457000.$}  &
  \multicolumn{1}{c}{[s]} &
  \multicolumn{1}{c}{[s]} &
  \multicolumn{1}{c}{model} &
  \multicolumn{1}{c}{residuals} &
  \multicolumn{1}{c}{model} &
  \multicolumn{1}{c}{residuals} \\
\hline\hline
  1331.91976 & $-$54.412& 0.154 & $-$56.976&  2.564&  $-$51.019& $-$3.392\\
  1345.90080 & $-$50.762& 0.149 & $-$53.998&  3.236&  $-$52.057&  1.295\\
  1415.83631 & $-$50.206& 0.170 & $-$49.765& $-$0.440&  $-$56.092&  5.886\\
  1430.69212 & $-$58.366& 0.151 & $-$51.050& $-$7.316&  $-$57.290& $-$1.076\\
  1497.33248 & $-$67.522& 0.159 & $-$64.881& $-$2.640&  $-$67.171& $-$0.350\\
  1510.38967 & $-$73.668& 0.153 & $-$68.905& $-$4.762&  $-$70.194& $-$3.474\\
  1523.23291 & $-$74.095& 0.156 & $-$73.198& $-$0.897&  $-$73.549& $-$0.546\\
  1538.50182 & $-$74.948& 0.194 & $-$78.679&  3.731&  $-$78.030&  3.081\\
  1576.10597 & $-$91.162& 0.159 & $-$93.498&  2.335&  $-$91.222&  0.059\\
  1589.91972 & $-$96.422& 0.150 & $-$99.251&  2.828&  $-$96.753&  0.330\\
  1604.82852 &$-$101.071& 0.186 &$-$105.545&  4.474& $-$103.061&  1.990\\
  1617.98105 &$-$109.424& 0.162 &$-$111.117&  1.692& $-$108.865& $-$0.558\\
  2041.24344 & $-$89.695& 0.160 & $-$79.666&$-$10.028&  $-$84.105& $-$5.589\\
  2053.84041 & $-$75.611& 0.158 & $-$70.408& $-$5.203&  $-$73.458& $-$2.153\\
  2066.69227 & $-$61.262& 0.139 & $-$60.657& $-$0.605&  $-$62.354&  1.091\\
  2079.92126 & $-$44.973& 0.164 & $-$50.340&  5.367&  $-$50.738&  5.764\\
  2150.46034 &  14.949& 0.143 &   7.536&  7.412&   11.516&  3.432\\
  2164.39670 &  26.042& 0.167 &  19.136&  6.906&   23.356&  2.685\\
  2235.00659 &  75.437& 0.157 &  75.446& $-$0.009&   77.893& $-$2.456\\
  2248.25195 &  82.000& 0.150 &  85.183& $-$3.182&   86.914& $-$4.913\\
  2261.29517 &  92.533& 0.152 &  94.410& $-$1.877&   95.404& $-$2.871\\
  2276.00083 & 101.932& 0.182 & 104.336& $-$2.404&   104.51& $-$2.586\\
  2287.40641 & 111.230& 0.157 & 111.655& $-$0.424&   111.26& $-$0.037\\
  2300.50082 & 119.058& 0.150 & 119.617& $-$0.558&   118.69&  0.362\\
  2313.64933 & 127.478& 0.159 & 127.108&  0.370&   125.83&  1.640\\
  2326.99985 & 135.568& 0.165 & 134.168&  1.400&   132.80&  2.762\\
  2340.08852 & 141.165& 0.144 & 140.530&  0.634&   139.40&  1.761\\
  2353.92851 & 146.208& 0.144 & 146.632& $-$0.423&   146.18&  0.020\\
\hline\end{tabular}\label{table:oc_v393car}
\tablefoot{The columns list the time stamp in $\textrm{BJD}_\textrm{TDB}$ \citep{Eastman2010} minus a constant, the $O-C$ value and its error $\sigma(O-C)$ as evaluated by the pipeline described in Section~\ref{Section:Harmonic_analysis}, the best-fit model with a linear baseline $(O-C)_\textrm{lin}$ (Eq.~\ref{eq:model_lin}) and its residuals, and the best-fit model with a quadratic baseline $(O-C)_\textrm{quad}$ (Eq.~\ref{eq:model_quad}) and its residuals. The unit of all columns except for the first one is seconds.}
\end{table*}

\section{Detected frequencies}
\begin{table}\centering
\caption{Significant GLS frequencies detected for Chang~134.}
\begin{tabular}{rlcl}
\hline
  \multicolumn{1}{c}{$\nu$ [c/d]}  &
  \multicolumn{1}{c}{$P$ [days]} &
  \multicolumn{1}{c}{GLS power} &
  notes\\
\hline\hline
   7.72651415 & 0.12942447 & 0.8961   & $\nu_0$ \\
  15.45302830 & 0.06471223 & 0.1057   & $2\nu_0$ \\
  23.17954244 & 0.04314149 & 0.0153   & $3\nu_0$ \\
  30.90605512 & 0.03235611 & 0.0019   & $4\nu_0$ \\
  38.63256381 & 0.02588489 & 4.48e-4  & $5\nu_0$ \\
  46.35909289 & 0.02157074 & 1.34e-4  & $6\nu_0$ \\ 
  54.08564236 & 0.01848919 & 4.35e-5  & $7\nu_0$ \\
  61.81213065 & 0.01617805 & 1.24e-5  & $8\nu_0$ \\
  69.53870051 & 0.01438048 & 3.68e-6  & $9\nu_0$ \\
\hline\end{tabular}\label{table:freq_chang134}
\tablefoot{The columns list the frequency $\nu$ in cycles per day, the corresponding period $P$ in days, the GLS power of the peak, and the identification of the mode or harmonics when available. The list is sorted by decreasing GLS power. See Section~\ref{Section:Harmonic_analysis} for details.}
\end{table}

\begin{table}\centering
\caption{Significant GLS frequencies detected for V393~Car.}
\begin{tabular}{rlcl}
\hline
  \multicolumn{1}{c}{$\nu$ [c/d]}  &
  \multicolumn{1}{c}{$P$ [days]} &
  \multicolumn{1}{c}{GLS power} & notes \\
\hline\hline
   7.07738170 & 0.14129519 & 0.9655  & $\nu_0$      \\    
  14.15476341 & 0.07064759 & 0.0275  & $2\nu_0$     \\
  21.23214517 & 0.04709840 & 0.0011  & $3\nu_0$     \\
  12.58288531 & 0.07947302 & 6.65e-4 & $\nu_1$      \\
  12.23921459 & 0.08170458 & 1.47e-4 & ---         \\
  11.85401165 & 0.08435962 & 1.22e-4 & ---         \\
  16.05643640 & 0.06228032 & 9.84e-5 & ---         \\
  25.28532371 & 0.03954863 & 9.33e-5 & $2\nu_1$?     \\
  13.57492256 & 0.07366524 & 8.19e-5 & ---         \\
  25.75951070 & 0.03882061 & 7.45e-5 & ---         \\
  19.66029065 & 0.05086394 & 4.37e-5 & $\nu_0+\nu_1$ \\
  18.79289737 & 0.05321159 & 2.68e-5 & ---         \\
  26.51035806 & 0.03772110 & 2.58e-5 & ---         \\
  18.20796573 & 0.05492101 & 2.06e-5 & ---         \\
  28.30950295 & 0.03532382 & 2.06e-5 & $4\nu_0$     \\
  20.85998893 & 0.04793866 & 1.99e-5 & ---         \\
  20.52339810 & 0.04872487 & 1.95e-5 & ---         \\
  24.12779695 & 0.04144597 & 1.32e-5 & ---         \\
  32.43606157 & 0.03082988 & 9.28e-6 & ---         \\
\hline\end{tabular}\label{table:freq_v393car}
\tablefoot{The columns list the frequency $\nu$ in cycles per day, the corresponding period $P$ in days, the GLS power of the peak, and the identification of the modes or harmonics when available. Here $\nu_1$ marks the secondary pulsation mode identified at $\sim$12.58 cycles per day, unlike Eq.~\ref{eq:freq_v393car}, where it marks the second most significant peak (corresponding to $2\nu_0$ in the present table). The list is sorted by decreasing GLS power. See Section~\ref{Section:Harmonic_analysis} for details.}
\end{table}

\end{document}